\documentclass[12pt,a4paper]{article}
\usepackage{color}
\usepackage{cite}
\usepackage{ulem}
\usepackage{latexsym,color}
\usepackage{epsfig}
\usepackage{amssymb,amsmath}

\newcommand{\bc}{\begin{center}}
\newcommand{\ec}{\end{center}}
\newcommand{\bd}{\begin{displaymath}}
\newcommand{\ed}{\end{displaymath}}
\newcommand{\be}{\begin{equation}}
\newcommand{\ee}{\end{equation}}
\newcommand{\ba}{\begin{array}}
\newcommand{\ea}{\end{array}}
\newcommand{\bt}{\begin{tabular}}
\newcommand{\et}{\end{tabular}}

\begin{document}

\title{Smallness of neutrino masses and leptogenesis in \\ 331 composite Higgs model
%Baryon asymmetry generation in the E$_6$CHM
}

\author{Roman Nevzorov\footnote{nevzorovrb@lebedev.ru}\\[5mm]
\itshape{I. E. Tamm Department of Theoretical Physics,}\\[0mm]
\itshape{Lebedev Physical Institute, Leninsky prospect 53, 119991 Moscow, Russia}}

\date{}

\maketitle

\begin{abstract}{
\noindent
We consider 331 composite Higgs model (CHM3) in which the Lagrangian of the strongly coupled sector is
invariant with respect to global $SU(3)_C \times SU(3)\times U(1)_6$ symmetry that can originate from
$SU(6)$ subgroup of $E_6$ and contains the gauge group of the standard model (SM) as a subgroup.
The breakdown of the approximate $SU(3)\times U(1)_6$ symmetry down to $SU(2)_W\times U(1)_Y$ subgroup
around the scale $f\sim 10\,\mbox{TeV}$ results in a set of pseudo--Nambu--Goldstone bosons (pNGBs) that,
in particular, involves Higgs doublet. The generation of the masses of the SM fermions in the CHM3 is
discussed. We argue that approximate $U(1)_L$ and discrete $Z_2$ symmetries may give rise to tiny masses
of the left--handed neutrinos and several composite fermions with masses $1-2\,\mbox{TeV}$.
The lepton and baryon asymmetries can be generated within the CHM3 via the out--of equilibrium decays of
extra Majorana particle into the Higgs doublet and these composite fermions.}
\end{abstract}

\newpage
\section{Introduction}
The smallness of three neutrino masses may be naturally explained within seesaw models \cite{see-saw}.
In these models the lepton asymmetry can be induced via the out--of equilibrium decays of the lightest right--handed
neutrino \cite{Fukugita:1986hr} since all three Sakharov conditions \cite{Sakharov:1967dj} are satisfied in this case.
This asymmetry gets partially converted into a baryon asymmetry via sphaleron processes \cite{Kuzmin:1985mm}.

In this context it is especially interesting to consider the possible origin of the mass hierarchy in the lepton sector
and leptogenesis within well motivated extensions of the standard model (SM) that allow to almost stabilize the
electroweak (EW) scale such as composite Higgs models (CHMs). These extensions of the SM proposed
in the 70's \cite{Terazawa:1976xx} and 80's \cite{composite-higgs} involve two sectors.
Apart from the weakly--coupled sector, that includes elementary states with the quantum numbers of
the SM fermions and SM gauge bosons, there are their composite partners which are formed in the second strongly
interacting sector \cite{Bellazzini:2014yua}. The Higgs doublet in these models is a set of bound states.

The Lagrangian of the strongly interacting sector of the minimal composite Higgs model (MCHM) \cite{Agashe:2004rs,Contino:2006qr}
possesses an approximate global $SU(3)_C \times SO(5)\times U(1)_X$ symmetry. $SU(3)_C\times SU(2)_W\times U(1)_Y$ gauge
group of the SM is a subgroup of $SU(3)_C \times SO(5)\times U(1)_X$. Near the scale $f$ the approximate
$SO(5)\times U(1)_X$ symmetry is expected to be broken down to $SO(4)\times U(1)_X \cong SU(2)_W\times SU(2)_R\times U(1)_X $.
Such breakdown results in four pseudo--Nambu--Goldstone bosons (pNGBs) that compose the Higgs doublet $H$.
The custodial symmetry $SU(2)_{cust} \subset SO(4) \cong SU(2)_W\times SU(2)_R$ \cite{Sikivie:1980hm} protects
the Peskin--Takeuchi $\hat{T}$ parameter \cite{Peskin:1991sw} against new physics contributions.
The contributions of new bound states to the EW observables in the CHMs were analysed in Refs. \cite{EWPOCHM}--\cite{Vignaroli:2012si}.
The implications of these models were also examined for Higgs physics \cite{Bellazzini:2012tv}--\cite{Azatov:2013ura}, \cite{Mrazek:2011iu}--\cite{Pomarol:2012qf}, gauge coupling unification \cite{Gherghetta:2004sq}--\cite{Barnard:2014tla},
dark matter \cite{Frigerio:2011zg}, \cite{Frigerio:2012uc}, \cite{Barnard:2014tla}--\cite{Asano:2014wra} and
collider phenomenology \cite{Pomarol:2008bh}--\cite{Bellazzini:2012tv}, \cite{Barbieri:2008zt},  \cite{Pomarol:2012qf}, \cite{Redi:2011zi}--\cite{Delaunay:2013pwa}. Different aspects of non-minimal CHMs were explored in Refs. \cite{Frigerio:2011zg}, \cite{Mrazek:2011iu}--\cite{Frigerio:2012uc}, \cite{Barnard:2014tla}--\cite{Asano:2014wra}, \cite{Cacciapaglia:2014uja}.

The SM bosons and fermions except the Higgs doublet are superpositions of elementary states and their composite partners.
Within the partial compositeness framework \cite{Contino:2006nn, Kaplan:1991dc} the couplings of the SM states
to the Higgs doublet are determined by the fractions of their compositeness.
The mass hierarchy in the quark and charged lepton sectors can be reproduced in this case if the fractions of compositeness
of all fermions except the top quark are rather small. This also leads to the partial suppression of flavour--changing processes \cite{Contino:2006nn}.
The non--diagonal flavour transitions in the quark and lepton sectors within the CHMs were
examined in Refs. \cite{Barbieri:2012tu}--\cite{Vignaroli:2012si}, \cite{Redi:2011zi},
\cite{Blanke:2008zb}--\cite{Barbieri:2012uh} and \cite{Redi:2013pga}, \cite{Barbieri:2012uh}--\cite{Csaki:2008qq}, respectively.
It was shown that in general the compositeness scale $f$  in these models is required to be larger than $10\,\mbox{TeV}$
\cite{Barbieri:2012tu}--\cite{Csaki:2008zd}, \cite{Redi:2011zi}, \cite{Blanke:2008zb}, \cite{Agashe:2006iy}.
However in the CHMs with extra flavour symmetries the phenomenologically acceptable scenarios
can be obtained even for $f\sim 1\,\mbox{TeV}$ \cite{Barbieri:2008zt}--\cite{Barbieri:2012tu}, \cite{Redi:2011zi}--\cite{Redi:2013pga},
\cite{Barbieri:2012uh}, \cite{Cacciapaglia:2007fw}.

In this note the mass hierarchy in the lepton sector and leptogenesis are discussed in the framework of 331 composite Higgs model (CHM3).
The strongly interacting sector of the CHM3 possesses a global $SU(3)_C \times SU(3)\times U(1)_6$ symmetry which contains
the SM gauge group as a subgroup and may stem from $SU(6)$ subgroup of $E_6$. The $E_6$ inspired composite Higgs models (E$_6$CHM) were
explored in Refs. \cite{Nevzorov:2015sha,Belyakova:2024fcw}\footnote{The $E_6$ inspired supersymmetric extensions of the SM were studied
in \cite{e6ssm}.}. Near the compositeness scale $f$ the breakdown of the approximate $SU(3)\times U(1)_6$
symmetry takes place giving rise to the composite Higgs doublet. Since the $SU(3)$ symmetry does not
contain $SU(2)_{cust}$ subgroup the electroweak precision measurements imply that $f\gtrsim 5-6\,\mbox{TeV}$ \cite{Nevzorov:2015sha}.

The paper is organised as follows.
In the next section we specify the CHM3, discuss the generation of masses of the SM fermions and argue
that approximate $U(1)_L$ and discrete $Z_2$ symmetries can lead to tiny masses of the left--handed neutrino.
The approximate $Z_2$ symmetry may also result in relatively light composite partners of the left--handed
leptons which might be detected at the LHC in the near future.
The leptogenesis in such scenario is considered in section 3. Our results are summarised in section 4.

\section{CHM3 with approximate $U(1)_L$ and $Z_2$ symmetries}

The measured Higgs mass $m_h\approx 125\,\mbox{GeV}$ corresponds to a quite small value of the Higgs quartic
coupling, $\lambda\simeq 0.13$ in the SM Higgs potential. On the other hand in general $\lambda$ tends to be
of the order of unity in the CHMs. The relatively small values of $m_h$ and $\lambda$ indicate that the Higgs doublet can
appear as a set of pNGB states from the breaking of an approximate global symmetry of the second strongly coupled sector.
The couplings of the elementary states to the fields of the strongly interacting sector explicitly break this global symmetry.
The pNGB potential arises from loops that contain elementary states. Therefore the quartic coupling $\lambda$ in the
Higgs effective potential is somewhat suppressed.

Within the CHMs it is convenient to present the pNGB states as
\begin{equation}
\Sigma= e^{i\Pi/f}\,,\qquad \Pi=\Pi^{\hat{a}} T^{\hat{a}}\,,
\label{1}
\end{equation}
where $T^{\hat{a}}$ are broken generators of the global symmetry. The breaking $SU(3)\times U(1)_6\to SU(2)_W\times U(1)_Y$
can be parameterised through the fundamental representation of $SU(3)$, i.e.
\begin{equation}
\Omega = \Sigma\, \Omega_0\,,\qquad\qquad \Omega_0^T = (0\quad 0\quad 1)\,.
\label{2}
\end{equation}
When $f$ is much larger than the vacuum expectation value of the Higgs field $v\simeq 246\,\mbox{GeV}$
the first two components of $\Omega$ can be identified with the SM--like Higgs doublet while the third
component involves the SM singlet scalar $\phi_0$.
The masses of all composite resonances except the Higgs doublet in the model under consideration
are set by the scale $f\gtrsim 5-6\,\mbox{TeV}$. For instance, the SM singlet scalar $\phi_0$
tends to acquire a mass in the multi TeV range. Thus all new states predicted by the CHM3 are rather
heavy and decouple from the SM spectrum.

In the CHM3 the $U(1)_Y$ weak hypercharges $Y_i$ are linear combinations
\begin{equation}
Y_i=\frac{2\sqrt{3}}{6} T^8_{i}+Q^6_i\,,
\label{3}
\end{equation}
where $Q^6_i$ are $U(1)_6$ charges of different multiplets and in the $SU(3)$ fundamental representation
$$
T^8=\dfrac{1}{2\sqrt{3}}\left(
\begin{array}{ccc}
1 & 0 & 0 \\
0 & 1 & 0 \\
0 & 0 & -2
\end{array}
\right)\,.
$$
To ensure the appropriate breakdown of $SU(3)\times U(1)_6$ symmetry $\Omega$ has to carry $U(1)_6$ charge (+1/3).

In the MCHM all composite partners of the SM fermions can be embedded into the vector representations of $SO(5)$ \cite{Contino:2006qr}, i.e.
${\bf 5}$--plets. This $SO(5)$ representation has the following decomposition in terms of $SU(2)_W\times SU(2)_R$:
\begin{equation}
{\bf 5}=({\bf 2, 2}) \oplus {\bf 1}\,.
\label{4}
\end{equation}
In Eq.~(\ref{4}) the first and second quantities in brackets are the $SU(2)_W$ and $SU(2)_R$ representations. In particular,
the composite partners of the left--handed and charged right--handed leptons ($L'_{1i}$ and $E^{'c}_{i}$) may belong to
${\bf 5}^i_{-1}$ and ${\bf 5}^i_{+1}$ respectively where lower index corresponds to the $U(1)_X$ charge of the {\bf 5}--plet
whereas $i=1,2,3$ runs over all three generations. However the interactions $(L'_{1i} H) (L'_{1j} H)$, that could generate
the masses of the left--handed neutrino, are forbidden in this model by the $U(1)_X$ symmetry. Such interactions may appear
if another composite partners $L'_{2i}$ of the left--handed leptons, which are components of ${\bf 5}^i_{0}$, are included.
The ${\bf 5}^i_{0}$ contain the following $SU(2)_W\times U(1)_Y$ multiplets
\begin{equation}
X_{i} = \left({\bf 2},\,+\dfrac{1}{2}\right)\,,\qquad\qquad L'_{2i} = \left({\bf 2},\,-\dfrac{1}{2}\right)\,,\qquad\qquad N_{i} = ({\bf 1},\, 0)\,.
\label{5}
\end{equation}
where the first and second numbers in brackets are the $SU(2)_W$ representation and $U(1)_Y$ charge.
The composite partners of the right--handed up-- and down--quarks ($U^{'c}_{i}$ and $D^{'c}_{i}$) are components of ${\bf 5}^i_{-2/3}$
and ${\bf 5}^i_{+1/3}$ respectively. There are two types of the partners of the left--handed quarks ($Q'_{1i}$ and $Q'_{2i}$)
in this case that belong to ${\bf 5}^i_{+2/3}$ and ${\bf 5}^i_{-1/3}$.

In the simplest $SU(5)$ Grand Unified Theories (GUTs) the left-handed quark and lepton doublets ($q_i$ and $\ell_i$),
the right-handed charged leptons ($e_i^c$), the right-handed up- and down-type quarks ($u_{i}^c$ and $d_i^c$) are components
of the $SU(5)$ multiplets, i.e.
\begin{equation}
u^c_{i} \in {\bf 10}_{i}\,,\qquad q_i \in {\bf 10}_i\,,\qquad d^c_i \in {\bf \overline{5}}_i\,,\qquad
e^c_i \in {\bf 10}_i\,,\qquad \ell_i \in {\bf \overline{5}}_i\,.
\label{6}
\end{equation}
The Lagrangian of the strongly interacting sector of the CHM3 is invariant under the transformations of the
global $SU(3)_C \times SU(3)\times U(1)_6$ symmetry which is a subgroup of $SU(6)$. The normalisation of the
$U(1)_6$ charges used in this paper implies that the fundamental representation of the $SU(6)$ group involves
\begin{equation}
{\bf 6} = \left({\bf 3},\,{\bf 1},\,-\dfrac{1}{3}\right) \oplus \left({\bf 1},\,{\bf 3},\,\dfrac{1}{3}\right)\,.
\label{7}
\end{equation}
Hereafter the first, second and third quantities in brackets are the $SU(3)_C$ and $SU(3)$ representations as well as $U(1)_{6}$ charges.
The ${\bf 5}$--plet and ${\bf 10}$--plet of $SU(5)$ can belong to ${\bf 6}$--plet, ${\bf 15}$--plet and ${\bf 20}$--plet of $SU(6)$.
These $SU(6)$ representations have the following decomposition in terms of $SU(5)$ representations:
${\bf 6}={\bf 5} \oplus {\bf 1}$, ${\bf 15}={\bf 10} \oplus {\bf 5}$ and ${\bf 20}={\bf 10} \oplus {\bf \overline{10}}$.

Here we assume that the composite partners of the right-handed up- and down-type quarks ($U_{i}^c$ and $D_i^c$) stem from
${\bf 20}$--plets and ${\bf \overline{15}}$--plets of $SU(6)$ so that they belong to
\begin{equation}
U^c_{i} \in \left({\bf \overline{3}},\,{\bf 3},\,-\dfrac{1}{3}\right)\,,\qquad
D^c_{i} \in \Biggl({\bf \overline{3}},\,{\bf \overline{3}},\,0\Biggr)\,.
\label{8}
\end{equation}
Then there are two types of the composite partners of the left--handed quarks ($Q_{1i}$ and $Q_{2i}$) which can be
components of ${\bf 15}$--plets and ${\bf 20}$--plets. Therefore we have
\begin{equation}
Q_{1i} \in \Biggl({\bf 3},\,{\bf 3},\,0\Biggr)\,,\qquad
Q_{2i} \in \left({\bf 3},\, {\bf \overline{3}},\, \dfrac{1}{3}\right)\,.
\label{9}
\end{equation}
Taking into account that the Higgs doublet is expected to be a part of the fundamental representation of $SU(6)$ (${\bf 6}_H$),
so that $\Omega = \left({\bf 1},\, {\bf 3},\, \dfrac{1}{3}\right)$, the generalisation of the $SU(5)$ structure of the quark Yukawa
interactions to the case of $SU(6)$ symmetry is given by \cite{Nevzorov:2015sha}
\begin{equation}
\mathcal{L}^q_{SU(6)}\sim Y^u_{ij}\, {\bf 20}(U^c_i) \times {\bf 15}(Q_{1j}) \times {\bf 6}_H +
Y^d_{ij}\, {\bf 20}(Q_{2i}) \times {\bf \overline{15}}(D^c_j) \times {\bf \overline{6}}_H + h.c. \,.
\label{10}
\end{equation}
The $SU(6)$ structure of interactions (\ref{10}) gives rise to the Yukawa couplings within the CHM3 that permit to induce the
masses of all SM quarks at low energies.

It is also expected that the composite partners of the right-handed charged leptons originate from ${\bf 15}$--plets
of $SU(6)$ whereas two types of the partners of the left--handed leptons come from ${\bf \overline{15}}$--plets
and ${\bf \overline{6}}$--plets. As a consequence one finds
\begin{equation}
\begin{array}{c}
E^c_{i} \in {\bf \overline{3}}^i_{+2/3} = \left({\bf 1},\,{\bf \overline{3}},\,\dfrac{2}{3}\right)\,,\qquad
L_{1i} \in {\bf 3}^i_{-2/3} = \left({\bf 1},\,{\bf 3},\,-\dfrac{2}{3}\right)\,,\\
L_{2i} \in {\bf \overline{3}}^i_{-1/3} = \left({\bf 1},\,{\bf \overline{3}},\,-\dfrac{1}{3}\right)\,.
\end{array}
\label{11}
\end{equation}
Multiplets ${\bf \overline{3}}^i_{-1/3}$ and ${\bf 3}^i_{+1/3}$ get combined forming vectorlike fermion states.
In order to ensure that the left--handed neutrino states gain masses, which are much smaller than the charged lepton masses,
approximate $U(1)_L$ and discrete $Z_2$ symmetries are imposed. All multiplets except ${\bf 3}^i_{+1/3}$ are even
under approximate $Z_2$ while ${\bf 3}^i_{+1/3}$ are $Z_2$ odd. The $U(1)_L$ symmetry is associated with the lepton
number conservation. Multiplets ${\bf \overline{3}}^i_{+2/3}$ and ${\bf 3}^i_{-2/3}$
carry $U(1)_L$ charges $(-1)$ and $(+1)$ respectively whereas all other multiplets of the composite states have zero $U(1)_L$ charges.
The Lagrangian of the strongly coupled sector respects the $U(1)_L$ global symmetry. Therefore the operators
${\bf \overline{3}}^i_{+2/3} \Omega^{\dagger} {\bf \overline{3}}^i_{-1/3}$ are forbidden and only interactions
\begin{equation}
\tilde{Y}^{e}_{ij} f ({\bf \overline{3}}^i_{+2/3} \Omega)(\Omega^{\dagger} {\bf 3}^j_{-2/3})
\label{12}
\end{equation}
permit to reproduce in the strongly coupled sector the Yukawa couplings
\begin{equation}
\mathcal{L}_{chl} \simeq Y^e_{ij} E^c_i (L_{1j} H^c) + h.c.\,,
\label{13}
\end{equation}
which give rise to non--zero masses of the charged leptons in the SM.

On the other hand the interactions $(L_{1i} H) (L_{1j} H)$ and $(L_{1i} H) (L_{2j} H)$, that could induce the masses
of the left--handed neutrino, are forbidden by the $U(1)_L$ symmetry. The appropriate interactions in the strongly
coupled sector
\begin{equation}
\mathcal{L}_{nl} \simeq \dfrac{\varkappa_{ij}}{f} (L_{2i} H)(L_{2j} H)
\label{14}
\end{equation}
may come from the operators
\begin{equation}
\tilde{\varkappa}_{ij} f ({\bf \overline{3}}^i_{-1/3} \Omega)({\bf \overline{3}}^j_{-1/3} \Omega)
\label{15}
\end{equation}
which are not suppressed by neither $U(1)_L$ nor $Z_2$ symmetry.
The $SU(6)$ structure, that leads to the interactions (\ref{12}) and (\ref{15}), can be written as
\begin{equation}
\mathcal{L}^e_{SU(6)}\sim \eta_{ij} \Biggl( {\bf 15}(E^c_i)\times {\bf \overline{6}}_H \Biggr) \times
\Biggl({\bf \overline{15}}(L_{1j})\times {\bf 6}_H \Biggr)
+ \xi_{ij} \Biggl( {\bf \overline{6}}(L_{2i})\times {\bf 6}_H \Biggr)\times \Biggl({\bf \overline{6}}(L_{2j})\times {\bf 6}_H\Biggr)\,.
\label{16}
\end{equation}

The mixing between the elementary lepton doublets and their composite partners results in the SM lepton doublets $\ell_i$
as well as vectorlike fermion doublets $\tilde{L}_{1i}$ and $\tilde{L}_{2i}$ so that
\begin{equation}
\begin{array}{rcl}
L_{1i} &=& s_{1i} \ell_i + c_{11i} \tilde{L}_{1i} + s_{12i} \tilde{L}_{2i}\,, \\
L_{2i} &=& s_{2i} \ell_i + c_{22i} \tilde{L}_{2i} + s_{21i} \tilde{L}_{1i}\,,
\end{array}
\label{17}
\end{equation}
where $c_{11i}\approx c_{22i}\approx 1$ while $s_{1i}$, $s_{2i}$, $s_{12i}$ and $s_{21i}$ are quite small.
Substituting $L_{2i}$ from Eqs.~(\ref{17}) into Eq.~(\ref{14}) one obtains the following
leading order analytic expression for the neutrino mass matrix ($\mathcal{M}^{\nu}_{ij}$)
$$
\mathcal{M}^{\nu}_{ij}= - \dfrac{\varkappa_{ij}}{2f} s_{2i} s_{2j} v^2\,.
$$
From Eqs.~(\ref{17}) and (\ref{13}) it follows that the masses of the charged leptons are determined
by $Y^e_{ij}\sim 1$, $s_{1i}$ and the fractions of the compositeness of the charged right--handed
leptons $s_{ei}$. All measured charged lepton masses
can be reproduced when $|s_{1i}|$ and $|s_{ei}|$ vary between $0.001$ and $0.1$\,.
Since $\varkappa_{ij}\sim 1$, the appropriate neutrino mass scale ($\sim 0.1\,\mbox{eV}$) requires
that $s_{2i}\lesssim 10^{-5}$, i.e. $s_{2i}\ll s_{1i}$.

The smallness of $s_{2i}$ can be caused by both $U(1)_L$ and approximate $Z_2$ symmetries.
If the mixing between $\tilde{L}_{2i}$ and elementary states vanishes then
$s_{12i}=s_{21i}=s_{2i}=0$ and the Lagrangian of the CHM3 possesses an exact global $U(1)_L$ symmetry
because in the strongly interacting sector the lepton number violating operators cannot be induced.
When $Z_2$ symmetry is exact, $\tilde{L}_{2i}$ remain massless and the values of $s_{12i}$, $s_{21i}$ and
$s_{2i}$ go to zero. The approximate $Z_2$ symmetry implies that $s_{2i}$ are small and
$\tilde{L}_{2i}$ are much lighter than $f$. The scenarios with relatively light
$\tilde{L}_{2i}$ are going to be considered in the next section.

%The smallness of $s_{2i}$ can be caused by both $U(1)_L$ and approximate $Z_2$ symmetries.
%In this case $\tilde{L}_{2i}$ can be also much lighter than $f$. The scenarios with relatively light
%$\tilde{L}_{2i}$ are going to be considered in the next section.

The approximate $Z_2$ symmetry may also lead to rather small values of $s_{2i}$ in the MCHM.
Nevertheless from Eq.~(\ref{5}) it follows that the mass terms of the components of ${\bf 5}^i_{0}$ are not suppressed
by the approximate $Z_2$ symmetry. As a consequence they can gain masses of the order of $f$.

\section{Leptogenesis in the CHM3}

Hereafter we explore the CHM3 with an approximate $Z_2$ symmetry that leads to three generations of relatively light composite
resonances $\tilde{L}_{2i}$. The SM singlet components $N_i$ of the corresponding $SU(3)$ multiplets ${\bf \overline{3}}^i_{-1/3}$
gain masses around $f$ through the interactions (\ref{15}) while the $SU(2)_W$ doublet components of these multiplets
are substantially lighter than $f$. This scenario is realized in Nature if the compositeness scale $f\gtrsim 10\,\mbox{TeV}$.
In this case the masses of $\tilde{L}_{2i}$ may vary between $1-2\,\mbox{TeV}$ so that it might be possible to discover
such fermion states in the near future at the LHC. The couplings of these composite resonances to the SM leptons are suppressed
because they are determined by the small parameters $s_{2i}$.

As mentioned before, if $f\gtrsim 10\,\mbox{TeV}$ one does not need to impose any extra flavour
symmetries to suppress the non--diagonal flavour transitions in the CHMs.
Here we assume that the low energy effective Lagrangian of the CHM3 is invariant with respect to a
$Z^B_2$ symmetry, which is a discrete subgroup of $U(1)_{B}$, i.e.
$$
Z^B_{2}=(-1)^{3B} \, ,
$$
where $B$ is the baryon number. This symmetry provides a mechanism for the suppression of
the operators which lead to rapid proton decay. Other processes that violate baryon number,
most notably neutron--antineutron transitions, are sufficiently strongly suppressed by the relatively
large value of $f\gtrsim 10\,\mbox{TeV}$ as well as the rather small mixing between
elementary states and their composite partners. Meanwhile for so large $f$ a substantial
degree of tuning is required to get a $125\,\mbox{GeV}$ Higgs state. Consequently, here
we focus on the scenarios with $f\simeq 10\,\mbox{TeV}$.

Now suppose that the weakly--coupled sector includes an additional elementary Majorana fermion which mixes
with the SM singlet components of ${\bf \overline{3}}^i_{-1/3}$. In general, since such mixing does not violate
neither $U(1)_L$ nor $Z_2$ symmetry, the compositeness fractions of the corresponding Majorana mass eigenstate $n$
and other SM fermions are expected to be of the same order of magnitude, i.e. $\sim |s_{1i}|\sim |s_{ei}|$.
If $n$ and $N_i$ have masses around $10\,\mbox{TeV}$ they can decay into leptons and Higgs doublet inducing lepton asymmetry.
We further assume that $N_i$ are somewhat heavier than the elementary fermion $n$. The part of the CHM3 Lagrangian, which describes
the interactions of $n=N_0$ and $N_i$ with the SM leptons, $\tilde{L}_{2i}$ and Higgs doublet $H$, can be presented as a sum
\begin{eqnarray}
\mathcal{L}_{N}=  g_{ix} \ell_i H N_x + h_{jx} \tilde{L}_{2j} H N_x + h.c.\,,
\label{18}
\end{eqnarray}
where $i,j=1,2,3$ and $x=0,1,2,3$. Because $\ell_i$ and $N_0$ are mainly elementary fields
$|g_{i0}|\ll |g_{ij}|\lesssim |h_{j0}| \ll |h_{ij}|$. At the same time the Yukawa couplings $h_{ij}$ are not suppressed
and therefore $|h_{ij}|\gtrsim 0.1$.

To simplify our analysis, we ignore the couplings $g_{ix}$, neglect the masses of $\tilde{L}_{2i}$ and set $h_{20}=h_{30}=0$.
Then the process of the lepton asymmetry generation is controlled by only one CP (decay) asymmetry
\begin{equation}
\varepsilon_{0}\simeq \dfrac{\Gamma_{L_{21}}-\Gamma_{\overline{L}_{21}}}
{\left(\Gamma_{L_{21}}+\Gamma_{\overline{L}_{21}}\right)}
\label{19}
\end{equation}
which appears on the right--hand side of Boltzmann equations.
Here $\Gamma_{L_{21}}$ and $\Gamma_{\overline{L}_{21}}$ are
partial decay widths of $n\to \tilde{L}_{21} + H$ and $n\to \overline{\tilde{L}}_{21} + H^{*}$.
At the tree level the decay asymmetry (\ref{19}) vanishes because
\begin{equation}
\Gamma_{L_{21}}=\Gamma_{\overline{L}_{21}}=\dfrac{|h_{10}|^2}{16 \pi}\,M_{n}\,.
\label{20}
\end{equation}
In Eq.~(\ref{20}) $M_{n}$ is the mass of the Majorana fermion $n$.
The CP violation gives rise to the non--zero value of $\varepsilon_{0}$ that
appears because of the interference between the tree--level amplitude of the decays of $n$
and one--loop corrections to it. For $h_{1x}=|h_{1x}| e^{i\varphi_{x}}$ and real values of $M_{n}$ and $M_{j}$
the calculation of one--loop diagrams gives \cite{Davidson:2008bu}
\begin{equation}
\varepsilon_{0}\simeq \dfrac{1}{8\pi}\Biggl[\sum_{j=1,2,3} |h_{1j}|^2 g\left(\frac{M^2_j}{M_n^2}\right) \sin 2\Delta\varphi_{j}\Biggr]\,,\qquad\qquad
\Delta\varphi_{j}=\varphi_{j}-\varphi_{0}\,,
\label{21}
\end{equation}
$$
g(x)=\sqrt{x}\biggl[\dfrac{1}{1-x}+1-(1+x)\ln\dfrac{1+x}{x}\biggr]\,,
$$
where $M_{j}$ are the masses of the composite resonances $N_j$. When CP invariance is preserved, i.e. $\varphi_{x}\to 0$,
the decay asymmetry (\ref{21}) vanishes. The absolute value of $\varepsilon_{0}$ reaches its maximum for
$\Delta\varphi_j=\Delta\varphi=\pm\pi/4$.

It is convenient to introduce $Y_{\Delta B}$ which is the baryon asymmetry relative to the entropy density.
The observed value of $Y_{\Delta B}$ is given by (see, for example, \cite{Davidson:2008bu})
\begin{equation}
Y_{\Delta B}=\dfrac{n_B-n_{\bar{B}}}{s}\biggl|_0=(8.75\pm 0.23)\times 10^{-11}\,.
\label{22}
\end{equation}
In order to compute $Y_{\Delta B}$ the system of Boltzmann equations, that permit to calculate the evolution of the number
density of $n$ and $(B-L)$ number asymmetry generated by the decays of $n$, must be solved.
The corresponding differential equations are somewhat similar to the Boltzmann equations describing thermal leptogenesis in the SM
which are specified in Refs.~\cite{Davidson:2008bu,Giudice:2003jh}. In the first approximation the induced baryon asymmetry
can be estimated using an approximate formula \cite{Davidson:2008bu,Giudice:2003jh}
\begin{equation}
Y_{\Delta B}\sim 10^{-3}\,\varepsilon_{0}\, \eta_{0}\,.
\label{23}
\end{equation}
In Eq.~(\ref{23}) $\eta_0$ is an efficiency factor
that varies from 0 to 1. Without washout effects the decays of the Majorana fermion $n$ would result
in $\eta_0=1$. The washout processes reduce the generated baryon asymmetry by the factor $\eta_0$.

For temperatures $T\lesssim 10\,\mbox{TeV}$, that correspond to the times $t\gtrsim 10^{-14}$ sec., the $(B+L)$
violating rate mediated by sphalerons is getting suppressed. Here we assume that the Majorana fermions $n$ decay
before the sphaleron processes shut off, i.e. $H(T=M_1)\lesssim {\Gamma}$, where $H$ is the Hubble expansion rate
and $\Gamma$ is the total decay rate of $n$. In this so--called strong washout scenario the efficiency
factor $\eta_0$ can be estimated as \cite{Giudice:2003jh}
\begin{eqnarray}
\eta_0 \simeq \left(\dfrac{H(T=M_1)}{2\Gamma}\right)^{1.16}\,,\qquad\qquad\nonumber\\
\Gamma = \Gamma_{L_{21}} + \Gamma_{\overline{L}_{21}}\,,
\qquad\qquad
H=1.66\, g_{*}^{1/2}\dfrac{T^2}{M_{Pl}}\,,
\label{24}
\end{eqnarray}
where $g_{*}=n_b+\dfrac{7}{8}\,n_f$ is the number of relativistic degrees of freedom in the thermal bath.
In the scenario under consideration $g_{*}=128.75$.

The requirement $H(T=M_1)\lesssim {\Gamma}$ implies that $|h_{10}|\gtrsim 10^{-6}$.
According to Eqs.~(24), for such values of $|h_{10}|$ the efficiency factor $\eta_0$
diminishes with increasing $|h_{10}|$. The corresponding dependence is shown in Fig.~1\,.
In Eq.~(\ref{23}) both the decay asymmetry $\varepsilon_{0}$ and the efficiency factor are smaller than unity. Thus the
appropriate baryon asymmetry can be generated only if $\eta_0\gtrsim 10^{-7}$. The results of the numerical analysis presented in Fig.~1
demonstrate that for $M_n\simeq 10\,\mbox{TeV}$ this condition can be fulfilled when $|h_{10}|\ll 0.001$. Otherwise
the absolute value of $Y_{\Delta B}$ tends to be too small.
The relatively small $|h_{10}|\lesssim 10^{-5}$ needed for successful leptogenesis
might be an indication that the Lagrangian of the CHM3 possesses an approximate $Z'_2$ symmetry,
under which all multiplets except $n$ are even while $n$ is $Z'_2$ odd.

%Within the CHM3 the value of $|h_{10}|$ is expected to be of the order of $s_{2i}$
%because fermion $n$ is mainly elementary state, i.e. $|h_{10}|\lesssim 10^{-5}$.

\setlength{\unitlength}{1cm}
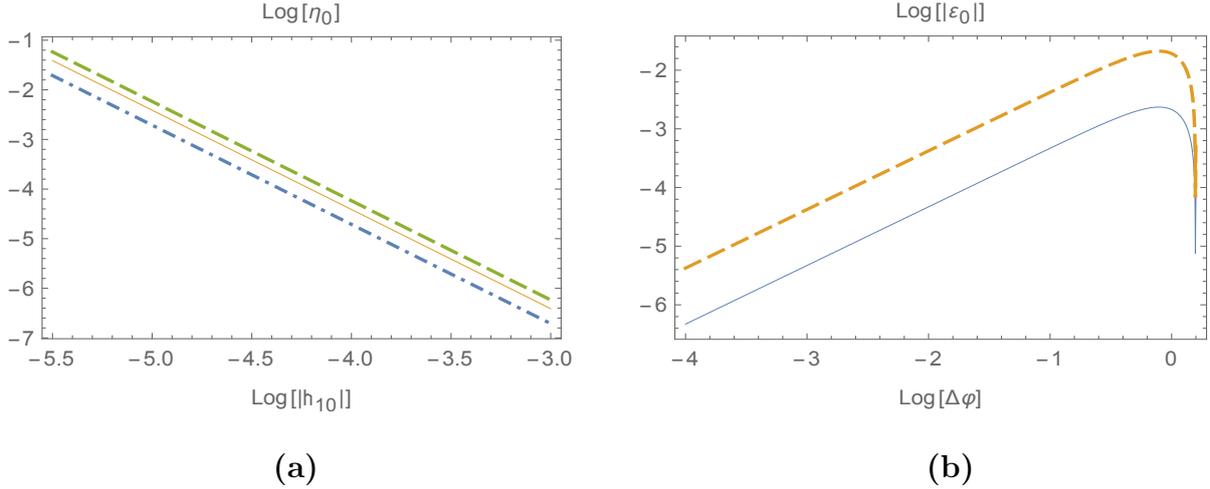
\begin{figure}[t]
\begin{picture}(16,8.5)
%\linethickness{1pt}
\put(0,3){\includegraphics[width=75mm,height=55mm]{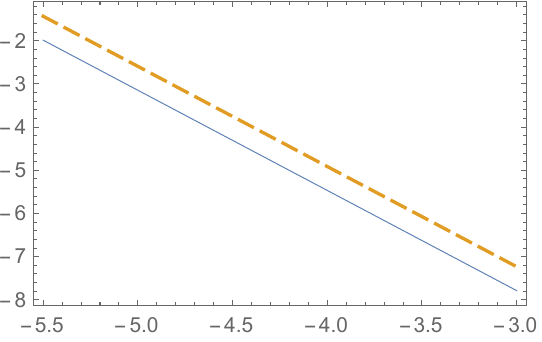}}
\put(0.1,8.8){$\log[|\eta_0|]$}
\put(3.2,2.7){$\log[|h_{10}|]$}
\put(8,3){\includegraphics[width=75mm,height=55mm]{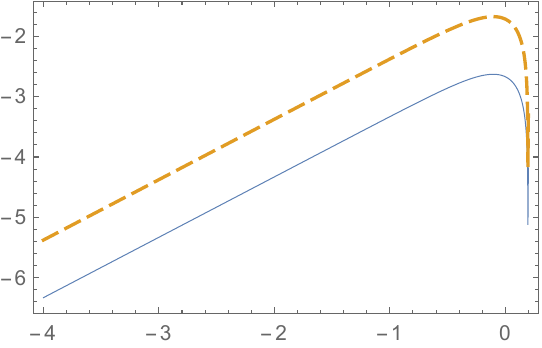}}
\put(8.1,8.8){$\log[|\varepsilon_{0}|]$}
\put(11.6,2.7){$\log[\Delta\varphi]$}
\end{picture}
\vspace{-3.5cm}
\caption{Logarithm (base 10) of the absolute values of the efficiency factor $\eta_0$ ({\bf Left}) and
decay asymmetry $\varepsilon_{0}$ ({\bf Right}) for $h_{20}=h_{30}=0$. The absolute value of $\eta_0$
is given as a function of logarithm (base 10) of $|h_{10}|$ for $M_{n}=10\,\mbox{TeV}$ (solid line) and
$M_{n}=15\,\mbox{TeV}$ (dashed line). The absolute value of $\varepsilon_{0}$ is presented as a function
of logarithm (base 10) of $\Delta\varphi_{1}=\Delta\varphi_{2}=\Delta\varphi_{3}=\Delta\varphi$ for
$M_{n}=10\,\mbox{TeV}$, $M_{1}=12\,\mbox{TeV}$, $M_{2}=14\,\mbox{TeV}$, $M_{3}=16\,\mbox{TeV}$,
$|h_{11}|=|h_{12}|=|h_{13}|=|h|$, $|h|=0.1$ (solid line) and $|h|=0.3$ (dashed line).}
\label{fig1}
\end{figure}

%\begin{figure}
%\includegraphics[width=75mm,height=55mm]{leptogen-chm3-1.pdf}\qquad
%\includegraphics[width=75mm,height=55mm]{leptogen-chm3-3.pdf}\\[2mm]
%\hspace*{3.5cm}{\bf (a)}\hspace*{8cm}{\bf (b) }\\
%\caption{Logarithm (base 10) of the absolute values of the efficiency factor $\eta_0$ and decay asymmetry $\varepsilon_{0}$
%for $h_{20}=h_{30}=0$. In {\it (a)} the absolute value of $\eta_0$ is given as a function of logarithm (base 10) of $|h_{10}|$
%for $M_{n}=5\,\mbox{TeV}$ (dashed--dotted line), $M_{n}=10\,\mbox{TeV}$ (solid line) and $M_{n}=15\,\mbox{TeV}$ (dashed line).
%In {\it (b)} the absolute value of $\varepsilon_{0}$ is presented as a function of logarithm (base 10) of
%$\Delta\varphi_{1}=\Delta\varphi_{2}=\Delta\varphi_{3}=\Delta\varphi$ for $M_{n}=10\,\mbox{TeV}$, $M_{1}=12\,\mbox{TeV}$,
%$M_{2}=14\,\mbox{TeV}$, $M_{3}=16\,\mbox{TeV}$, $|h_{11}|=|h_{12}|=|h_{13}|=|h|$, $|h|=0.1$ (solid line) and $|h|=0.3$ (dashed line).}
%\label{fig1}
%\end{figure}

Since in the scenario under consideration the efficiency factor may not be negligibly small, i.e. $\eta_0\sim 0.01$,
the absolute value of $Y_{\Delta B}$ is determined by the decay asymmetry $\varepsilon_{0}$.
From Eq.~(\ref{21}) one can see that $\varepsilon_{0}$ is set by $|h_{1j}|$ and CP violating phases $\Delta\varphi_{j}$.
It also depends on the ratio of masses of Majorana fermions $M_j/M_n$. Because the Yukawa couplings of $N_j$ to
$L_{2i}$ and the Higgs doublet are not suppressed, all $|h_{ij}|$ should be relatively
large, i.e. $|h_{ij}| \gtrsim 0.1$. Here we fix $M_{n}=10\,\mbox{TeV}$, $M_{1}=12\,\mbox{TeV}$,
$M_{2}=14\,\mbox{TeV}$, $M_{3}=16\,\mbox{TeV}$, $|h_{11}|=|h_{12}|=|h_{13}|=h$ and
$\Delta\varphi_{1}=\Delta\varphi_{2}=\Delta\varphi_{3}=\Delta\varphi$. In Fig.~1 the dependence of $|\varepsilon_{0}|$ on
$\Delta\varphi$ for different values of $h$ is shown. The absolute value of the decay asymmetry grows monotonically
with increasing of $|h|$. The results presented in Fig.~1 indicate that for $h\gtrsim 0.1$ and $\eta_0\sim 0.01$ the
observed baryon asymmetry can be obtained for small CP violating phases, i.e. $\Delta \varphi\ll 0.01$.

\section{Conclusions}

In this paper we have discussed the 331 composite Higgs model (CHM3) in which the strongly interacting sector possesses
approximate $SU(3)_C \times SU(3)\times U(1)_6$ symmetry. This global symmetry may originate from $SU(6)$ subgroup of $E_6$.
It is expected that near the scale $f\sim 10\,\mbox{TeV}$ the approximate $SU(3)\times U(1)_6$ symmetry is broken down
to the $SU(2)_W\times U(1)_Y$ subgroup giving rise to five pNGBs. These pNGB states form Higgs doublet and one SM singlet scalar.
The composite partners of the SM quarks and SM leptons belong to fundamental and antifundamental representations of $SU(3)$ with
different $U(1)_6$ charges.

We argued that the observed mass hierarchy in the lepton sector can be caused by the approximate
$U(1)_L$ and discrete $Z_2$ symmetries. Such $Z_2$ symmetry may also lead to three generations of relatively light composite neutral fermions
and fermions with charges $\pm 1$. These resonances compose $SU(2)_W$ doublets $\tilde{L}_{2i}$ and can have masses within the
$1-2\,\mbox{TeV}$ range so that they can be discovered at the LHC in the near future.
The couplings of $\tilde{L}_{2i}$ to the SM leptons are quite suppressed because they are defined by the small parameters $s_{2i}\lesssim 10^{-5}$.
It was shown that in this case the observed baryon asymmetry can be induced if the particle spectrum of the CHM3 contains
an additional elementary Majorana fermion $n$ with mass around $10\,\mbox{TeV}$. The corresponding lepton asymmetry may be generated due to
the out--of equilibrium decays of $n$ into $\tilde{L}_{2i}$ and Higgs doublet even when CP is almost preserved.

\section*{Acknowledgements}
RN is grateful to M.~G.~Belyakova, E.~E.~Boos, S.~V.~Demidov, D.~S.~Gorbunov, D.~I.~Kazakov, N.~V.~Krasnikov, D.~G.~Levkov and M.I.~Vysotsky
for fruitful discussions.

%Although the new scalar particle discovered at the LHC in 2012 is consistent with the Standard Model (SM) Higgs boson,
%it could in principle be composed of more fundamental degrees of freedom. The idea of a composite Higgs boson was
%proposed in the 70's \cite{Terazawa:1976xx} and 80's \cite{composite-higgs}. It implies the presence of a
%strongly interacting sector in which electroweak (EW) symmetry breaking (EWSB) is generated dynamically.
%Generically, in models of this type the composite Higgs tends to have a large quartic coupling $\lambda\gtrsim 1$.
%At the same time, the observed SM-like Higgs boson is relatively light and corresponds to $\lambda\simeq 0.13$.
%This indicates that the discovered Higgs state could possibly be a  pseudo-Nambu-Goldstone boson (pNGB)
%originating from the spontaneous breakdown of an approximate global symmetry of the strongly interacting sector.

%\newpage

\end{document}